\documentclass[11pt]{article}
\usepackage{a4,bm,epsfig,booktabs,stmaryrd,diagrams}
\usepackage{setspace}
\usepackage{ifthen,array}
\newcommand{\ams}{\usepackage{amsfonts,amssymb,amsmath}}

\ams
\allowdisplaybreaks[3]
\newlength{\textwidthorig}
\newlength{\oddsidemarginorig}
\newlength{\textheightorig}
\newlength{\topmarginorig}
\def\seitenlaengenabsolut#1 #2 #3 #4 {\setlength{\textwidth}{#1}
                                      \setlength{\oddsidemargin}{#2}
                                      \setlength{\textheight}{#3}
                                      \setlength{\topmargin}{#4}}
\def\seitenlaengenrelzustandard#1 #2 #3 #4 {\setlength{\textwidth}{\textwidthorig+#1}
                                            \setlength{\oddsidemargin}{\oddsidemarginorig+#2}
                                            \setlength{\textheight}{\textheightorig+#3}
                                            \setlength{\topmargin}{\topmarginorig+#4}}
\def\seitenlaengenrelzuvorher#1 #2 #3 #4 {\addtolength{\textwidth}{#1}
                                          \addtolength{\oddsidemargin}{#2}
                                          \addtolength{\textheight}{#3}
                                          \addtolength{\topmargin}{#4}}
\newcommand{\standardseite}{\seitenlaengenrelzuvorher2.2cm -0.8cm 1.8cm -1.5cm }   %
\standardseite

\newlength{\laengespatium}

\newcommand{\nach}{\longrightarrow}      

\newcommand{\auf}{\longmapsto}           
\newcommand{\txtauf}[1]{\auf}            
\newcommand{\impliz}{\Longrightarrow}    
 
\newcommand{\invimpliz}{\Longleftarrow}  
\newcommand{\iso}{\cong}                 
\newcommand{\ident}{\equiv}              
\newcommand{\teilmenge}{\subseteq}       
\newcommand{\obermenge}{\supseteq}       
\newcommand{\aeqrel}{\sim}               

\newcommand{\kreuz}{\times}              

\newcommand{\keil}{\wedge}               

\newcommand{\betraganpass}[1]%
           {\left| #1 \right|}           
\newcommand{\bigbetrag}[1]%
           {\bigl|{#1}\bigr|}            
\newcommand{\betrag}[1]%
           {|{#1}|}                      
\newcommand{\betragnichtanpass}[1]%
           {\mid #1 \mid}                
\newcommand{\norm}[1]%
           {{}{\parallel}#1{\parallel}{}}      
\newcommand{\erww}[1]%
           {\langle #1 \rangle}          
\newcommand{\skalprod}[2]%
           {\langle #1,#2 \rangle}       
\newcommand{\quer}{\overline}            
\newcommand{\inv}[1]{\frac{1}{#1}}       
\newcommand{\del}{\partial}                            
\newcommand{\dd}{\text{d}}                             
\newcommand{\I}{\text{i}}                              
\newcommand{\EINS}{{\boldsymbol{1}}}                   
\newcommand{\field}[1]{\mathbb{#1}}                    
\newcommand{\C}{{\field{C}}}                           
\newcommand{\R}{{\field{R}}}                           
\newcommand{\Gl}{\text{Gl}}                            
\newcommand{\gl}{\mathfrak{gl}}                        
\newcommand{\Sl}{\text{Sl}}                            
\newcommand{\rnkl}[2]{\raisebox{-0.4ex}{$#1$}%
\raisebox{-0.12ex}{{\large$\setminus$}}\,#2}   
\newcommand{\agb}{{\overline{{\cal A}/{\cal G}}}}      
\newcommand{\agbfact}[1][]{\text{$\agb/\!\aeqrel$}}    
\newcommand{\Gb}{{\overline{{\cal G}}}}                

\newcommand{\qa}{{\quer{A}}}                           

\newcommand{\holgr}{{\mathbf H}}                       
\newcommand{\bz}{{\mathbf B}}                          

\newcommand{\LG}{{\mathbf{G}}}                         
\newcommand{\aeqrelzush}[1][]{\sim}                    
\newcommand{\alg}{\mathfrak{A}}                          
\newcommand{\blg}{\mathfrak{B}}                          
\newcommand{\malg}{{\cal M}(\alg)}                     

\newcommand{\nklza}[1][]{\ifthenelse{\equal{#1}{}}     
                                    {\rnkl{Z(\holgr_\qa)}{\LG}}        
                                   {\rnkl{Z(\holgr_{#1})}{\LG}}}       
\newcommand{\nkla}[1][]{\ifthenelse{\equal{#1}{}}      
                                    {\rnkl{\bz(\qa)}{\Gb}}        
                                    {\rnkl{\bz(#1)}{\Gb}}}       

\newcommand{\charakt}{\chi}                            

\newcommand{\YM}{{\text{YM}}}                          

\newcommand{\ymwirk}[1][]{\ifthenelse{\equal{#1}{}}{S_{\YM}}{S_{\YM,#1}}}

\newcommand{\bmat}{\begin{pmatrix}}
\newcommand{\emat}{\end{pmatrix}}

\newcommand{\ListNullAbstaende}{\setlength{\topsep}{0pt}%
                                \setlength{\parskip}{0pt}%
                                \setlength{\partopsep}{0pt}%
                                \setlength{\itemsep}{0pt}%
                                \setlength{\parsep}{0pt}}
\newcommand{\ListNurAnstrichAbstand}{\setlength{\topsep}{0pt}%
                                     \setlength{\parskip}{0pt}%
                                     \setlength{\partopsep}{0pt}%
                                     \setlength{\parsep}{0pt}}
\newenvironment{StandardListe}[2]%
               {\begin{list}%
                      {#1}%
                      {\settowidth{\leftmargin}{M#1}%
                       \settowidth{\labelwidth}{#1}%
                       \settowidth{\labelsep}{M}%
                       #2%
                      }%
                }%
               {\end{list}}%
\newenvironment{EinfachListe}[1]%
               {\begin{StandardListe}{#1}{\ListNullAbstaende}}%
               {\end{StandardListe}}%
               {\begin{StandardListe}{#1}{\ListNurAnstrichAbstand}}%
               {\end{StandardListe}}%
\newcommand{\labelsatz}[1]{#1}
\newcounter{listennr}                      %
\newlength{\hilfslaenge}
\newlength{\stdlabellaenge}
\newlength{\maximum}
\newcommand{\stdlabel}{}
\newcommand{\Maximum}{}
\newcommand{\iitem}[1][]{\ifthenelse{\equal{#1}{}}%
                           {\item \setlength{\hilfslaenge}{\stdlabellaenge}}%
                           {\item[\labelsatz{#1}\hfill]%
                            \settowidth{\hilfslaenge}{\labelsatz{#1}}}%
                         \ifthenelse{\lengthtest{\maximum < \hilfslaenge}}%
                           {\setlength{\maximum}{\hilfslaenge}%
                            \ifthenelse{\equal{#1}{}}%
                               {\renewcommand{\Maximum}{\stdlabel}}%
                               {\renewcommand{\Maximum}{#1}}}%
                           {}%
                      }      
\makeatletter
\newenvironment{AutoLabelLaengenListe}[2][]%
               {\begin{list}%
                      {\labelsatz{#1}\hfill}%
                      {\stepcounter{listennr}%
                       \settowidth{\leftmargin}{M\labelsatz{\ref{listnr\arabic{listennr}}}}%
                       \settowidth{\labelwidth}{\labelsatz{\ref{listnr\arabic{listennr}}}}%
                       \settowidth{\labelsep}{M}%
                       \settowidth{\stdlabellaenge}{\labelsatz{#1}}%
                       \renewcommand{\stdlabel}{#1}%
                       #2%
                       \renewcommand{\Maximum}{}%
                      }%
                }%
               {\renewcommand{\@currentlabel}{\Maximum}%
                \label{listnr\arabic{listennr}}%
                \end{list}%
                }%
\makeatother
\newenvironment{StandardEinrueckung}[2]%
               {\begin{list}%
                      {#1}%
                      {\settowidth{\leftmargin}{M#1}%
                       \settowidth{\labelwidth}{#1}%
                       \settowidth{\labelsep}{M}%
                       #2%
                      }%
                \item}%
               {\end{list}}%
\newenvironment{Einrueckungpur}[1]%
               {\begin{StandardEinrueckung}{#1}{\ListNullAbstaende}}%
               {\end{StandardEinrueckung}}%
\newenvironment{Einrueckung}[1]%
               {\begin{StandardEinrueckung}{#1}{\setlength{\parsep}{0pt}}}%
               {\end{StandardEinrueckung}}%

\makeatletter
\newcommand{\EineNumZeileGleichung}[2][0.5ex]
           {
            
            \vspace{#1} 
            \noindent
            \stepcounter{equation}
            \renewcommand{\@currentlabel}{\arabic{equation}}%
            \phantom{(\arabic{equation})}\hspace*{\fill}
            $\displaystyle{#2}$
            \hspace*{\fill}
            (\arabic{equation})

            \vspace{#1} 
            
           }
\makeatother
\makeatletter
\newcommand{\EineErwNumZeileGleichung}[2][0.5ex]
           {
            
            \vspace{#1} 
            \noindent
            \stepcounter{equation}
            \renewcommand{\@currentlabel}{\arabic{equation}}%
            \phantom{(\arabic{equation})}\hspace*{\fill}
            #2 %
            \hspace*{\fill}
            (\arabic{equation})

            \vspace{#1} 
            
           }
\makeatother
\newcommand{\breitrel}[1]{\hspace*{\tabcolsep} #1 \hspace*{\tabcolsep}}
\newlength{\abstaug}              %
\newenvironment{AllgUnnumGleichung}[2][1.0ex]
               {
  
                \setlength{\abstaug}{#1}
                \vspace{\abstaug}
                \hspace*{\fill}
                $\begin{array}[t]{#2}
                }%
               {\end{array}$
                \hspace*{\fill}
  
                \vspace{\abstaug}

                }%
\newenvironment{AllgNumGleichung}[2][0.0ex]
               {
  
                \setlength{\abstaug}{#1}
                \vspace{\abstaug}
                $\begin{tabular*}{\textwidth}[t]{#2}
                }%
               {\end{tabular*}$

                \vspace{\abstaug}

               }%
\newenvironment{StandardUnnumGleichungKlein}[1][0ex]
               {\renewcommand{\s}{\\[#1] }%
                \begin{AllgUnnumGleichung}{rcl}}%
               {\end{AllgUnnumGleichung}}%
\newcommand{\s}{\\[0ex] }             %
\newenvironment{StandardUnnumGleichung}[1][0ex]%
               {\renewcommand{\s}{\\[#1] }%
                \begin{AllgUnnumGleichung}{>{\displaystyle}rc>{\displaystyle}l}}%
               {\end{AllgUnnumGleichung}}%
\newenvironment{XrelYZNumGleichung}[1][0ex]
               {\renewcommand{\s}{\\[#1] }%
                \begin{AllgNumGleichung}{rcll}}%
               {\end{AllgNumGleichung}}%
\newcommand{\erl}[1]{\hfill\mbox{\hspace*{1.5em}\small (#1)}}

\newcommand{\erllang}[2][0.5\textwidth]%
              {\hfill\hspace*{1.5em}%
               \begin{minipage}[t]{#1}{\small%
                          \begin{list}{(}{\ListNullAbstaende%
                                          \settowidth{\leftmargin}{(}%
                                          \settowidth{\labelwidth}{(}%
                                          \settowidth{\labelsep}{}%
                                         }%
                          \item#2)%
                          \end{list}}%
               \end{minipage}\\[-0.9ex]
              }%
\newcommand{\DefBemUmgeb}[1]%
           {\newenvironment{#1}[1][]%
                           {\begin{Einrueckung}{{\bf #1}}%
                            \ifx##1\empty\else{{\bf ##1}
                            
                                                        }\fi%
                            }%
                           {\end{Einrueckung}}}
\newcommand{\DefSBemUmgeb}[2]
           {\newenvironment{#1}[1][]%
                           {\begin{Einrueckung}{{\bf #2}}%
                            \ifx##1\empty\else{{\bf ##1}
                            
                                                        }\fi%
                            }%
                           {\end{Einrueckung}}}
\makeatletter
\newcommand{\DefBspUmgeb}[3]
           {\newcounter{#2}[#3]%
            \newenvironment{#1}[1][]%
                           {\stepcounter{#2}%
                            \renewcommand{\ZaehlerMarke}{\arabic{#2}}%
                            \renewcommand{\Einzugsname}{{\bf #1 \ZaehlerMarke}}%
                            \begin{Einrueckung}{\Einzugsname}
                            \ifx##1\empty\else{{\bf ##1}\\}\fi%
                            \renewcommand{\@currentlabel}{\ZaehlerMarke}%
                            }%
                           {\end{Einrueckung}}}
\makeatother
\newcommand{\ZaehlerbisEbene}{section}
\newcommand{\Ebenea}{section}
\newcommand{\Ebeneb}{subsection}

\newcommand{\Abschnittnummer}{%
            \ifx\ZaehlerbisEbene\Ebenea{\arabic{section}}%
             \else{%
              \ifx\ZaehlerbisEbene\Ebeneb{\arabic{section}.\arabic{subsection}}%
               \else{\arabic{section}.\arabic{subsection}.\arabic{subsubsection}}%
              \fi}%
            \fi}     
\newcommand{\Einzugsname}{}
\newcommand{\ZaehlerMarke}{}
\makeatletter
\newcommand{\DefThmUmgeb}[3]%
           {\newcounter{#1}[#3]%
            \newenvironment{#1}[1][]%
                           {\stepcounter{#2}%
                            \setcounter{#1}{\value{#2}}%
                            \renewcommand{\ZaehlerMarke}{\Abschnittnummerpunkt\arabic{#1}}%
                            \renewcommand{\Einzugsname}{{\bf #1 \ZaehlerMarke}}%
                            \begin{Einrueckung}{\Einzugsname}
                            \ifx##1\empty\else{{\bf ##1}
                            
                                                        }\fi%
                            \renewcommand{\@currentlabel}{\ZaehlerMarke}%
                            }%
                           {\end{Einrueckung}}}
\makeatother
\makeatletter
\newcommand{\DefSThmUmgeb}[4]%
           {\newcounter{#1}[#3]%
            \newenvironment{#1}[1][]%
                           {\stepcounter{#2}%
                            \setcounter{#1}{\value{#2}}%
                            \renewcommand{\ZaehlerMarke}{\Abschnittnummerpunkt\arabic{#1}}%
                            \renewcommand{\Einzugsname}{{\bf #4 \ZaehlerMarke}}
                            \begin{Einrueckung}{\Einzugsname}
                            \ifx##1\empty\else{{\bf ##1}

                                                        }\fi%
                            \renewcommand{\@currentlabel}{\ZaehlerMarke}%
                            }%
                           {\end{Einrueckung}}}
\makeatother
\makeatletter
\newcommand{\DefUnterNumThmUmgeb}[5]%
           {\newcounter{#1}[#3]%
            \newcounter{#4}%
            \newenvironment{#1}[1][]%
                           {\ifx##1\empty\else{\stepcounter{#2}\setcounter{#4}{0}}\fi%
                            \stepcounter{#4}%
                            \setcounter{#1}{\value{#2}}%
                            \renewcommand{\ZaehlerMarke}{\Abschnittnummerpunkt\arabic{#1}\alph{#4}}%
                            \renewcommand{\Einzugsname}{{\bf #5 \ZaehlerMarke}}
                            \begin{Einrueckung}{\Einzugsname}
                            \renewcommand{\@currentlabel}{\ZaehlerMarke}%
                            }%
                           {\end{Einrueckung}}}
\makeatother
\newenvironment{Beweis}[1][]%
               {\begin{Einrueckung}{{\bf Beweis}}%
                \ifx#1\empty\else{{\bf #1}

                                            }\fi%
                }%
               {\end{Einrueckung}%
                }%
\newenvironment{Proof}[1][]%
               {\begin{Einrueckung}{{\bf Proof}}%
                \ifx#1\empty\else{{\bf #1}

                                            }\fi%
                }%
               {\end{Einrueckung}%
                }%
               {\begin{Einrueckung}{{\bf \glqq Beweis\grqq}}%
                \ifx#1\empty\else{{\bf #1}
                
                                            }\fi%
                }%
               {\end{Einrueckung}%
                }%
               {\begin{Einrueckung}{{\bf Begr"undung}}%
                \ifx#1\empty\else{{\bf #1}
                
                                            }\fi%
                }%
               {\end{Einrueckung}%
                }%
\newenvironment{Hinrichtung}%
               {\begin{Einrueckungpur}{$\impliz$}}%
               {\end{Einrueckungpur}}%
\newenvironment{Rueckrichtung}%
               {\begin{Einrueckungpur}{$\invimpliz$}}%
               {\end{Einrueckungpur}}%
               {\begin{Einrueckungpur}{\glqq$\teilmenge$\grqq}}%
               {\end{Einrueckungpur}}%
               {\begin{Einrueckungpur}{\glqq$\obermenge$\grqq}}%
               {\end{Einrueckungpur}}%
               {\begin{Einrueckungpur}{"$\teilmenge$"}}%
               {\end{Einrueckungpur}}%
               {\begin{Einrueckungpur}{"$\obermenge$"}}%
               {\end{Einrueckungpur}}%
\newcommand{\qed}{\nopagebreak\hspace*{2em}\hspace*{\fill}{\bf qed}}
\newcommand{\ARabic}{\arabic}
\newcommand{\Nummerntypa}{\arabic}   
\newcommand{\Nummerntypb}{\alph}
\newcommand{\Nummerntypc}{\roman}
\newcommand{\Nummerntypd}{\Alph}

\newcommand{\Nra}{\Nummerntypa{Nummera}}            %
\newcommand{\Nrb}{\Nummerntypb{Nummerb}}            %
\newcommand{\Nrc}{\Nummerntypc{Nummerc}}                
\newcommand{\Nrd}{\Nummerntypd{Nummerd}}                
\newcommand{\ZeichenzuNrTyp}[1]%
           {\ifx#1\ARabic {.}\else{)}%
                  \fi}                              %
\newcommand{\NrZeicha}{\ZeichenzuNrTyp{\Nummerntypa}}
\newcommand{\NrZeichb}{\ZeichenzuNrTyp{\Nummerntypb}}
\newcommand{\NrZeichc}{\ZeichenzuNrTyp{\Nummerntypc}}
\newcommand{\NrZeichd}{\ZeichenzuNrTyp{\Nummerntypd}}
\newcommand{\ListMarkea}%
           {\Nra\NrZeicha}
\newcommand{\ListMarkeb}%
           {\Nra\NrZeicha\Nrb\NrZeichb}
\newcommand{\ListMarkec}%
           {\Nra\NrZeicha\Nrb\NrZeichb\Nrc\NrZeichc}
\newcommand{\ListMarked}%
           {\Nra\NrZeicha\Nrb\NrZeichb\Nrc\NrZeichc\Nrd\NrZeichd}
\newcommand{\Anfangszeichen}{}
\newcommand{\Anfangspunkt}{}
\newcounter{Schachtelebene}
\newcounter{Hilfszaehler}
\newcommand{\Hilfsbefehl}{}
\newcommand{\Schachtelebene}{\alph{Schachtelebene}}
\makeatletter
\newenvironment{AllgNumerierteListe}[2][]
               {\addtocounter{Schachtelebene}{1}%
		\setcounter{Hilfszaehler}{#2}%
                \renewcommand{\Anfangszeichen}%
                             {\renewcommand{\Hilfsbefehl}{\csname Nummerntyp\Schachtelebene \endcsname}%
                              \Hilfsbefehl{Hilfszaehler}}%
                \renewcommand{\Anfangspunkt}%
                             {\csname NrZeich\Schachtelebene \endcsname}%
                \begin{list}%
                      {\stepcounter{Nummer\Schachtelebene}%
                       \csname Nr\Schachtelebene \endcsname
                       \csname NrZeich\Schachtelebene \endcsname
                       }%
                      {\settowidth{\leftmargin}{M\Anfangszeichen\Anfangspunkt}%
                       \settowidth{\labelwidth}{\Anfangszeichen\Anfangspunkt}%
                       \settowidth{\labelsep}{M}%
                       \setlength{\topsep}{0pt}%
                       \setlength{\parskip}{0pt}%
                       \setlength{\partopsep}{0pt}%
                       \setlength{\itemsep}{0pt}%
                       \setlength{\parsep}{0pt}%
                      }%
                \renewcommand{\@currentlabel}{\csname ListMarke\Schachtelebene \endcsname}%
                }%
               {\ifthenelse{\equal{}{}}{\setcounter{Nummer\Schachtelebene}{0}}{}
                \addtocounter{Schachtelebene}{-1}%
                \end{list}}
\makeatother
\newenvironment{NumerierteListe}[1]
               {\begin{AllgNumerierteListe}{#1}}
               {\end{AllgNumerierteListe}}
\newenvironment{WeiterNumerierteListe}[1]
               {\begin{AllgNumerierteListe}[Weiter]{#1}}
               {\end{AllgNumerierteListe}}

\newcommand{\UnnumAnfangszeichen}{}
\newcounter{UnnumSchachtelebene}
\newcommand{\UnnumSchachtelebene}{\alph{UnnumSchachtelebene}}
\makeatletter
\newenvironment{UnnumerierteListe}%
               {\addtocounter{UnnumSchachtelebene}{1}%
                \renewcommand{\UnnumAnfangszeichen}%
                             {\csname UnnumZeich\UnnumSchachtelebene \endcsname}%
                \begin{list}%
                      {\UnnumAnfangszeichen}%
                      {\settowidth{\leftmargin}{M\UnnumAnfangszeichen}%
                       \settowidth{\labelwidth}{\UnnumAnfangszeichen}%
                       \settowidth{\labelsep}{M}%
                       \setlength{\topsep}{0pt}%
                       \setlength{\parskip}{0pt}%
                       \setlength{\partopsep}{0pt}%
                       \setlength{\itemsep}{0pt}%
                       \setlength{\parsep}{0pt}%
                      }%
                }%
               {\addtocounter{UnnumSchachtelebene}{-1}%
                \end{list}}
\makeatother
\newlength{\fktdefhilfslaenge}
\newcommand{\ohnefktdef}[4]
           {\hspace*{\fill}
            $\begin{array}[t]{ccc}%
            #1 & \nach & #2 \\
            #3 & \auf  & #4
            \end{array}$
            \hspace*{\fill}}
\newcommand{\fktdef}[5]
           {\hspace*{\fill}
            $\begin{array}[t]{cccc}%
            #1: & #2 & \nach & #3 \\    
                & #4 & \auf  & #5
            \end{array}$
            \settowidth{\fktdefhilfslaenge}{$#1$:}
            \hspace*{0.6 \fktdefhilfslaenge}  
            \hspace*{\fill}}
\newcommand{\fktdefpur}[5]
           {$\begin{array}[t]{cccc}%
            #1: & #2 & \nach & #3 \\    
                & #4 & \auf  & #5
            \end{array}$}
\newcommand{\fktdefabgesetztpur}[5]
           {
            
            $\begin{array}[t]{cccc}%
            #1: & #2 & \nach & #3 \\    
                & #4 & \auf  & #5
            \end{array}$
            \settowidth{\fktdefhilfslaenge}{$#1$:}
            \hspace*{0.6 \fktdefhilfslaenge}
            
           }
\newcommand{\fktdefabgesetzt}[5]
           {
           
            \hspace*{\fill}
            $\begin{array}[t]{cccc}%
            #1: & #2 & \nach & #3 \\    
                & #4 & \auf  & #5
            \end{array}$
            \settowidth{\fktdefhilfslaenge}{$#1$:}
            \hspace*{0.6 \fktdefhilfslaenge}  
            \hspace*{\fill}
            
            }
\newcommand{\ohnefktdefabgesetzt}[4]
           {      

            \hspace*{\fill}
            $\begin{array}[t]{ccc}%
            #1 & \nach & #2 \\
            #3 & \auf  & #4
            \end{array}$
            \hspace*{\fill}

            }
\newcommand{\doppelohnefktdefabgesetzt}[6]
           {

            \hspace*{\fill}
            $\begin{array}[t]{ccccc}%
            #1 & \nach & #2 & \nach & #3\\
            #4 & \auf  & #5 & \auf  & #6
            \end{array}$
            \hspace*{\fill}

            }
\newcommand{\anhang}%
           {\appendix
            \sectioninh{Anhang}
            \renewcommand{\Abschnittnummer}{%
                  \ifx\ZaehlerbisEbene\Ebenea{\Alph{section}}%
                  \else{%
                        \ifx\ZaehlerbisEbene\Ebeneb{\Alph{section}.\arabic{subsection}}%
                        \else{\Alph{section}.\arabic{subsection}.\arabic{subsubsection}}%
                        \fi}%
                  \fi}%
                 
            }            
\newcommand{\anhangengl}%
           {\appendix
            \sectioninh{Appendix}
            \renewcommand{\Abschnittnummer}{%
                  \ifx\ZaehlerbisEbene\Ebenea{\Alph{section}}%
                  \else{%
                        \ifx\ZaehlerbisEbene\Ebeneb{\Alph{section}.\arabic{subsection}}%
                        \else{\Alph{section}.\arabic{subsection}.\arabic{subsubsection}}%
                        \fi}%
                  \fi}%
                 
            }

\newcounter{wdhlstufe}
\newcommand{\sectioninh}[1]%
           {\section*{#1}%
            \addcontentsline{toc}{section}{#1}}
\newcommand{\bezeichnung}[3]%
           {\begin{Einrueckungpur}{\hbox to 6em{#1}\hbox to 2.4em{\hfill#2}}
            #3
            \end{Einrueckungpur}}

\newcommand{\doppelteinfach}{e}

\newcommand{\ifdoppelt}[1]{\ifthenelse{\equal{\doppelteinfach}{d}}{#1}{}}
\newcommand{\ifeinfach}[1]{\ifthenelse{\equal{\doppelteinfach}{e}}{#1}{}}

\newlength{\querfhilfsl}              %

\newlength{\hll}

\DefThmUmgeb{Theorem}{Theorem}{\ZaehlerbisEbene}
\DefThmUmgeb{Definition}{Definition}{\ZaehlerbisEbene}
\DefThmUmgeb{Satz}{Theorem}{\ZaehlerbisEbene}
\DefThmUmgeb{Proposition}{Theorem}{\ZaehlerbisEbene}
\DefThmUmgeb{Lemma}{Theorem}{\ZaehlerbisEbene}
\DefThmUmgeb{Folgerung}{Theorem}{\ZaehlerbisEbene}
\DefThmUmgeb{Corollary}{Theorem}{\ZaehlerbisEbene}
\DefThmUmgeb{Vorschrift}{Definition}{\ZaehlerbisEbene}
\DefThmUmgeb{Construction}{Definition}{\ZaehlerbisEbene}
\DefSThmUmgeb{FormSatz}{Theorem}{\ZaehlerbisEbene}{\glqq Satz\grqq} 
\DefThmUmgeb{Vermutung}{Theorem}{\ZaehlerbisEbene}
\DefThmUmgeb{Conjecture}{Theorem}{\ZaehlerbisEbene}
\DefThmUmgeb{Konvention}{Definition}{\ZaehlerbisEbene}
\DefThmUmgeb{Feststellung}{Theorem}{\ZaehlerbisEbene}
\DefUnterNumThmUmgeb{DefinitionZusatzNum}{Definition}{\ZaehlerbisEbene}{DefZN}{Definition}
\DefBspUmgeb{Beispiel}{Beispiel}{subsubsection}
\DefBspUmgeb{Example}{Example}{subsubsection}
\DefBspUmgeb{Frage}{Frage}{section}
\DefBspUmgeb{Question}{Question}{section}
\DefBspUmgeb{Aufgabe}{Aufgabe}{section}
\DefBspUmgeb{Ziel}{Ziel}{section}
\DefBemUmgeb{Bemerkung}
\DefBemUmgeb{Remark}
\DefSBemUmgeb{OffeneFrage}{Offene Frage}
\newcommand{\bdf}{\begin{Definition}}
\newcommand{\edf}{\end{Definition}}
\newcommand{\bvorsch}{\begin{Vorschrift}}
\newcommand{\evorsch}{\end{Vorschrift}}
\newcommand{\bconst}{\begin{Construction}}
\newcommand{\econst}{\end{Construction}}
\newcommand{\bthm}{\begin{Theorem}}
\newcommand{\ethm}{\end{Theorem}}
\newcommand{\bsatz}{\begin{Satz}}
\newcommand{\esatz}{\end{Satz}}
\newcommand{\bprop}{\begin{Proposition}}
\newcommand{\eprop}{\end{Proposition}}
\newcommand{\blem}{\begin{Lemma}}
\newcommand{\elem}{\end{Lemma}}
\newcommand{\bfolg}{\begin{Folgerung}}
\newcommand{\efolg}{\end{Folgerung}}
\newcommand{\bcorr}{\begin{Corollary}}
\newcommand{\ecorr}{\end{Corollary}}
\newcommand{\bfest}{\begin{Feststellung}}
\newcommand{\efest}{\end{Feststellung}}
\newcommand{\bbew}{\begin{Beweis}}
\newcommand{\ebew}{\end{Beweis}}
\newcommand{\bpf}{\begin{Proof}}
\newcommand{\epf}{\end{Proof}}
\newcommand{\bwnum}{\begin{WeiterNumerierteListe}}
\newcommand{\ewnum}{\end{WeiterNumerierteListe}}
\newcommand{\bdfzn}{\begin{DefinitionZusatzNum}}
\newcommand{\edfzn}{\end{DefinitionZusatzNum}}
\newcommand{\bbem}{\begin{Bemerkung}}
\newcommand{\ebem}{\end{Bemerkung}}
\newcommand{\brem}{\begin{Remark}}
\newcommand{\erem}{\end{Remark}}
\newcommand{\bnum}{\begin{NumerierteListe}}
\newcommand{\enum}{\end{NumerierteListe}}
\newcommand{\bunum}{\begin{UnnumerierteListe}}
\newcommand{\eunum}{\end{UnnumerierteListe}}
\newcommand{\bbsp}{\begin{Beispiel}}
\newcommand{\ebsp}{\end{Beispiel}}
\newcommand{\bex}{\begin{Example}}
\newcommand{\eex}{\end{Example}}
\newcommand{\bfrag}{\begin{Frage}}
\newcommand{\efrag}{\end{Frage}}
\newcommand{\bquest}{\begin{Question}}
\newcommand{\equest}{\end{Question}}
\newcommand{\baufg}{\begin{Aufgabe}}
\newcommand{\eaufg}{\end{Aufgabe}}
\newcommand{\bof}{\begin{OffeneFrage}}
\newcommand{\eof}{\end{OffeneFrage}}
\newcommand{\bverm}{\begin{Vermutung}}
\newcommand{\everm}{\end{Vermutung}}
\newcommand{\bconj}{\begin{Conjecture}}
\newcommand{\econj}{\end{Conjecture}}
\newcommand{\bkonv}{\begin{Konvention}}
\newcommand{\ekonv}{\end{Konvention}}
\newcommand{\bglklein}{\begin{StandardUnnumGleichungKlein}}
\newcommand{\eglklein}{\end{StandardUnnumGleichungKlein}}
\newcommand{\bgl}{\begin{StandardUnnumGleichung}}
\newcommand{\egl}{\end{StandardUnnumGleichung}}
\newcommand{\bglrtext}{\begin{XrelYZNumGleichung}}
\newcommand{\eglrtext}{\end{XrelYZNumGleichung}}

\newcommand{\berlgl}{\begin{StandardUnnumGleichung}}
\newcommand{\eerlgl}{\end{StandardUnnumGleichung}}
\newcommand{\beinrueck}{\begin{Einrueckungpur}} 
\newcommand{\eeinrueck}{\end{Einrueckungpur}}
\newcommand{\beinflist}{\begin{EinfachListe}} 
\newcommand{\eeinflist}{\end{EinfachListe}}
\newcommand{\beq}{\begin{equation}}
\newcommand{\eeq}{\end{equation}}
\newcommand{\bhin}{\begin{Hinrichtung}}
\newcommand{\ehin}{\end{Hinrichtung}}
\newcommand{\brueck}{\begin{Rueckrichtung}}
\newcommand{\erueck}{\end{Rueckrichtung}}
\newcommand{\bvl}{\begin{AutoLabelLaengenListe}{\ListNullAbstaende}}
\newcommand{\evl}{\end{AutoLabelLaengenListe}}
\newcommand{\df}[1]{{\bf #1}}

\newlength{\adressabstand}
\diagramstyle[Postscript=dvips]
\newarrow{auf}|---{->}
\newarrow{nach}----{->}
\newarrow{ident}33333
\newarrow{aufsurj}|---{->>}
\newarrow{nachsurj}----{->>}
\newarrow{einbett}C---{->}
\newarrow{strichel}{}{dash}{}{dash}{->}
\newarrow{impliz}===={=>}
\newlength{\CDhoehe}                  
\newlength{\CDgap}                    
\newcommand{\CDstdlaenge}{3.5em}

\newcommand{\stdskalprod}[2]{\skalprod{#1}{#2}_{\mathrm{Eucl}}}

\newcommand{\chart}{\chi}
\newcommand{\spinbdl}{S}

\newcommand{\coveroben}{\Lambda}
\newcommand{\covergroup}{\lambda}
\newcommand{\kosmkonst}{\Lambda}
\newcommand{\hodge}{\ast}
\newcommand{\action}{\rho}
\newcommand{\weight}{k}
\newcommand{\metrik}{q}
\newcommand{\drei}{n}
\newcommand{\viermetrik}{g}
\newcommand{\hubble}{h}
\renewcommand{\Gl}{\mathrm{Gl}}
\renewcommand{\gl}{{\mathfrak{gl}}}
\renewcommand{\Sl}{\mathrm{Sl}}

\newcommand{\SO}{\mathrm{SO}}
\newcommand{\so}{{\mathfrak{so}}}
\newcommand{\Ocf}{\mathrm{O}}

\newcommand{\Spin}{\mathrm{Spin}}
\newcommand{\SU}{\mathrm{SU}}
\newcommand{\su}{{\mathfrak{su}}}
\newcommand{\extcurv}{K}
\newcommand{\bdlextcurv}{\widetilde\extcurv}
\newcommand{\wgabb}{W}
\newcommand{\ivp}[3][]{#2\bullet_{#1}#3}
\newcommand{\ivpalone}{\bullet}
\newcommand{\ashkov}{\nabla^A}
\newcommand{\laz}{A}
\newcommand{\conn}{\omega}
\newcommand{\spinconn}{\widetilde\omega_{LC}}
\newcommand{\lcconn}{\omega_{LC}}
\newcommand{\ashconn}{\omega_A}
\newcommand{\ashconnspin}{\widetilde\omega_A}

\newcommand{\viernabla}{{}^4\nabla}
\newcommand{\stdbasis}{\mathfrak e}
\newcommand{\stdx}{\mathfrak x}
\newcommand{\stdy}{\mathfrak y}
\newcommand{\base}{e}
\newcommand{\anotherbase}{f}
\newcommand{\invbase}{\varepsilon}
\newcommand{\invanotherbase}{\eta}
\newcommand{\cs}{\Sigma}   
\newcommand{\neueseite}{\newpage}
\newcommand{\nonicht}[1]{\textbf{Das ist hier noch unbesetzt.}}

\makeatletter
\newcounter{tab}
\newcommand{\fusszeile}{}
                        {\par\noindent\fusszeile
                         \end{center}}
\makeatother
\DefThmUmgeb{Convention}{Definition}{\ZaehlerbisEbene}

\newcommand{\Bigbetrag}[1]%
           {\Bigl|{#1}\Bigr|}            
\newcommand{\Biggbetrag}[1]%
           {\Biggl|{#1}\Biggr|}

\newcommand{\bconv}{\begin{Convention}}
\newcommand{\econv}{\end{Convention}}

\newcommand{\bpm}{\begin{pmatrix}}
\newcommand{\epm}{\end{pmatrix}}

\DeclareMathOperator{\spec}{spec}

\begin{document}
\title{Ashtekar Variables: Structures in Bundles} 
\author{Christian Fleischhack$^{1,2,\ast}$ and Philipp Levermann$^{2,3,}$\thanks{e-mail: 
             {\tt fleischh@math.upb.de}, {\tt karl.philipp.levermann@desy.de}} \\   
        \\
        {\normalsize\em $^{1}$Institut f\"ur Mathematik}\\[\adressabstand]
        {\normalsize\em Universit\"at Paderborn}\\[\adressabstand]
        {\normalsize\em Warburger Stra\ss e 100}\\[\adressabstand]
        {\normalsize\em 33098 Paderborn, Germany}
        \\[-25\adressabstand]      
        {\normalsize\em $^{2}$Department Mathematik}$^{\phantom{1}}$\\[\adressabstand]
        {\normalsize\em Universit\"at Hamburg}\\[\adressabstand]
        {\normalsize\em Bundesstra\ss e 55}\\[\adressabstand]
        {\normalsize\em 20146 Hamburg, Germany}
        \\[-25\adressabstand]
        {\normalsize\em $^{3}$II. Institut f\"ur Theoretische Physik}$^{\phantom{1}}$\\[\adressabstand]
        {\normalsize\em Universit\"at Hamburg}\\[\adressabstand]
        {\normalsize\em Luruper Chaussee 149}\\[\adressabstand]
        {\normalsize\em 22761 Hamburg, Germany}
        \\[-25\adressabstand]}     
\date{December 6, 2011}
\maketitle
\newcommand{\iotarestr}{\iota_\stdrestr}
\newcommand{\stdrestr}{\setabb}
\newcommand{\algrestr}[1][\stdrestr]{\alg_{#1}}
\newcommand{\blgrestr}[1][\stdrestr]{\blg_{#1}}
\newcommand{\clg}{{\mathfrak C}}
\newcommand{\dlg}{{\mathfrak D}}
\newcommand{\set}{\mathbf S}
\newcommand{\topset}{\set}   
\newcommand{\elset}{s} 
\renewcommand{\malg}{\spec \alg} 
\newcommand{\Y}{\mathbf{Y}} 
\newcommand{\ely}{\mathbf{y}} 
\newcommand{\restr}[1]{#1_{\stdrestr}}
\newcommand{\elalg}{a} 
\newcommand{\elblg}{b} 
\newcommand{\elclg}{c} 
\newcommand{\eldlg}{d} 
\newcommand{\mclg}{\spec \clg} 
\newcommand{\charabb}{\tau}
\newcommand{\charact}{\charakt}
\newcommand{\disjunion}{\sqcup}
\newcommand{\setabb}{\sigma}
\newcommand{\plus}{p}
\newcommand{\cover}{{\cal U}}
\newcommand{\linf}{\ell^\infty}
\newcommand{\gelf}[1][]{\ifthenelse{\equal{#1}{}}{\widetilde}{G_{#1}}}
\newcommand{\gelftrf}{{\sim}}
\newcommand{\coverbasis}{\cover_{\text{Basis}}}
\newcommand{\lin}{{\text{lin}}}
\newcommand{\bip}{\beta}

\begin{abstract}
Canonical gravity can be formulated by means of a densitized dreibein
together with an $\SU(2)$ connection. These so-called Ashtekar variables 
are the fundamental quantities, loop quantum gravity is resting on.
In this paper we review these variables from the perspective of 
fibre bundles. This is straightforward for the dreibein field as
this is simply a frame. The Ashtekar connection, however, is more complicated.
It turns out, that at the level of the tangent bundle,
it is given by the Levi-Civita connection plus a multiple of the Weingarten mapping,
whose action on vector fields is induced from the 
vector product on $\R^3$. Lifted to the spin bundle, one 
regains the well-known $\SU(2)$ Ashtekar connection.
At the end, we apply our results to FRW spacetimes.
\end{abstract}


\section{Introduction}
25 years ago, Abhay Ashtekar \cite{a128,a117}  
introduced new variables to describe canonical
gravity. Their main advantage has been that they drastically simplified the constraints
of gravity: These become polynomial. This opened a completely new door to approach
the quantization of gravity, ultimately leading to loop quantum gravity.
Over the last quarter of a century, there has been only a single, but significant
extension of Ashtekar's variables. 
In the mid-90s, Barbero \cite{e80} and Immirzi \cite{e81} 
added a new parameter $\bip$, with $\bip = \I$ giving the original variables.
It turned now out a big advantage that for real $\bip$ the structure group
is no longer $\Sl_\C(2)$, but $\SU(2)$. This has been crucial for the integration
theory of loop quantum gravity. Before the Barbero-Immirzi idea has been introduced,
complicated reality condition had been necessary to implement the real
structure of the theory. However, the new formulation has the drawback that
the Hamiltonian constraint is no longer polynomial. This has only been cured by the
Thiemann trick \cite{e1}, 
rewriting the term with prefactor $1+\bip^2$ by means of certain Poisson brackets.
That combined the advantages of integration theory on compact structure groups and
the functional analysis of polynomial constraints.

Despite the fundamental r\^ole of Ashtekar's variables, their geometric origin
have remained open. To the best of our knowledge, only local versions using 
sophisticated index notations have been available so far. 
The present paper aims at a first glimpse
of the differential geometry underlying Ashtekar's variables. 
These are a connection in some principal fibre bundle to be determined and
a densitized dreibein field. The latter one is rather easy to state, whence
we will focus on the connection variables which form the configuration space
of the theory (up to gauge transformations). More precisely, we will describe
the principal fibre bundle the connection lives in, and then discuss 
Ashtekar-type connections. Additionally, we present a very simple formulation
of these variables in the context of Friedmann-Robertson-Walker models.
In a forthcoming paper we will focus on the description of constraints.

The present paper is organized as follows: We start in Section \ref{sect:canongrav}
with a short review of canonical gravity mostly to fix the notations in 
Section \ref{sect:notation}, 
and comment, in Section \ref{sect:frames}, briefly on 
densitized frame fields being one part of Ashtekar's variables. The rest of this
paper is devoted to the Ashtekar connection.
Sloppily, on tangent space level, 
this connection is the sum of the Levi-Civita connection and
the Weingarten mapping (augmented by the Barbero-Immirzi parameter), the latter
one to be described in Subsection \ref{subsect:weingarten}. However, to make this
construction precise, we need to transfer the $\R^3$ vector product to a 
certain product on the tangent space (of a 3D manifold) to be given
in Subsection \ref{subsect:modif-kreuz-prod}. This leads to the Ashtekar $\SO(3)$
connection in Subsection \ref{subsect:frame-bundle-conn}. 
Introducing spin structures in Section \ref{sect:spin}, we get 
the usual Ashtekar $\SU(2)$ connection. 
Finally, we will apply the whole construction to homogeneous 
isotropic spacetimes in Section \ref{sect:cosmol}.

\section{Canonical Gravity}
\label{sect:canongrav}
Classically, gravity
\cite{Wald,Straumann}
is described within Einstein's theory of general relativity by a 
smooth metric $\viermetrik$ of Lorentzian signature on a $4$-di\-men\-sion\-al
manifold satisfying Einstein's equations
\bgl
R_{\mu\nu} - \inv2 \viermetrik_{\mu\nu} R + g_{\mu\nu} \kosmkonst
 & = & \frac{8\pi\gamma}{c^4} T_{\mu\nu} \,.
\egl\noindent
Here, $R_{\mu\nu}$ is the Ricci tensor w.r.t.\ $\viermetrik$, and $R$ the corresponding Ricci scalar.
$\kosmkonst$ denotes the cosmological constant. The right-hand side describes
the matter content of the universe; we will assume this term to vanish 
throughout our paper (vacuum case). 
We neglect $\kosmkonst$ as well.
In this paper we will restrict ourselves to a particular class of such manifolds --
that of so-called spacetimes. 
\bdf
A \df{spacetime} is a 4-dimensional, connected, time-oriented and oriented 
Lorentz manifold.
\edf
\noindent
Often, physical theories are quantized in their canonical form as this admits an
initial-value formulation. 
A key feature of classical general relativity is, however, the unified treatment 
of space and time -- or, rephrased, the lack of a standard splitting into 
space and time. Thus, instead of having a ``canonical'' initial hypersurface
at $t = 0$,
any spacelike hypersurface might serve as initial hypersurface.

For this, let us
assume we are given some Lorentzian metric $\viermetrik$
on $M$ fulfilling the vacuum Einstein equations, and some
spacelike hypersurface $\cs$ embedded into
$M$. The geometry of $\cs$ is determined by the Riemannian metric $\metrik$
induced by $\viermetrik$ and by the second fundamental form $\extcurv$.
The latter one is a symmetric bilinear form on $\cs$ and 
describes the shaping of $\cs$ within the spacetime $M$;
it is also called ``extrinsic curvature'' in physics.
However, $\metrik$ and $\extcurv$ cannot be any metric and any form.
In fact, they are induced from a metric on $M$ fulfilling the Einstein equations.
Therefore, the Gau\ss-Codazzi equations impose 
constraints on $(\metrik,\extcurv)$.

It has been a big issue whether 
these constraints are not only necessary, but also sufficient
for a Cauchy evolution. In her celebrated paper \cite{grav5},
Choquet-Bruhat showed that indeed the Cauchy problem of general relativity
has locally a unique solution.
The constraints are
sufficient to locally reconstruct the spacetime uniquely up to isometric isomorphisms
from a Riemannian metric $\metrik$ and a symmetric bilinear form $\extcurv$
on some 3-man\-i\-fold $\cs$.\ \cite{Straumann,Ringstrom}
Only globally, the situation is a bit more subtle. 
Usually, these problems can be circumvented by some additional physical assumptions.
To avoid, in particular, multiple intersections with timelike curves,
one has restricted oneself to a special type of 
spacetimes -- so-called globally hyperbolic ones.
This means, by definition \cite{grav1}, that
$M$ fulfills the strong
causality condition and for all $p,q \in M$ the intersection of the causal future of
$p$ and the causal past of $q$ is compact.
Indeed, Geroch \cite{grav7} has shown that any globally hyperbolic spacetime
has a Cauchy slice, i.e., some spacelike hypersurface of $M$ that is
intersected exactly once by each inextendible timelike curve in $M$ \cite{grav1};
moreover then $M \iso \R \kreuz \cs$.
However, Geroch was only able to prove this in the $C^0$ category.
It is still a rather recent result, that this statement is also true
in the smooth category.

\bthm
If a spacetime $(M,g)$ is globally hyperbolic, then it is isometric to
\bgl
(\R \kreuz \Sigma, -f \: \dd\tau^2 + g_\tau)
\egl
with a smooth positive function
$f : \R \nach \R$ and a smooth family of Riemannian metrics $g_\tau$ on $\Sigma$.
Moreover, each $\{t\} \kreuz \Sigma$ is a Cauchy slice.\
\cite{grav1,grav6}
\ethm

\section{Notation}
\label{sect:notation}

Unless specified otherwise,
let $\cs$ be some oriented three-dimensional manifold, and let $\metrik$ 
be some Riemannian metric $\metrik$ on $\cs$. 
We write $\skalprod X Y := \metrik(X,Y)$ for all vector fields $X,Y$ on $\cs$.
As $\cs$ is orientable, 
it has a spin structure and is parallelizable.\ \cite{Kirby}
Next, we assume that $M$ is some manifold diffeomorphic
to $\cs \kreuz \R$, whereas we identify $\cs$ with $\cs \kreuz \{0\}$.
Moreover, $\viermetrik$ is a metric on $M$ inducing
$\metrik$ on $\cs$.
We will denote $\viermetrik(X,Y)$ again by $\skalprod X Y$; this does not apply to
Section \ref{sect:cosmol}. 
Finally, observe that we are always working with smooth objects only.

\section{Frames}
\label{sect:frames}
One part of the Ashtekar variables is formed by a densitized 
dreibein, where dreibeine are orthonormal frames on a three-dimensional 
manifold $\cs$. In contrast to the Ashtekar connections to be defined later,
the definition of frames works in any dimension $\drei$. 

\subsection{General Frames}
\label{subsect:general-frames}

\bdf
A \df{frame at $x \in \cs$}
is a vector space isomorphism 
\bgl
\base : \R^\drei \nach T_x \cs \,.
\egl
\edf\noindent
Another way to specify a frame 
is to select
a basis of $T_x \cs$.\ \cite{KoNo1}
Recall that the frame bundle on $\cs$ is given by the 
disjoint union, indexed by $x \in \cs$, of all frames at $x$. 
The differentiable structure on it is naturally induced 
from that on $\cs$ by decomposing each frame w.r.t.\
some appropriate local coordinate system on $T\cs$.
The resulting bundle $\Gl(\cs)$ is independent of such a choice and 
called \df{frame bundle}. Its 
structure group is given by $\Gl(\drei)$. 
Frames are simply (local) sections in the frame bundle.

If we choose some (local) basis for $T_x\cs$ (and the canonical
basis of $\R^\drei$), then any frame $\base$ at $x$ as being a vector 
space isomorphism, is characterized by some matrix. 
Its determinant will be called 
\df{determinant} $\det e$ of $\base$.
If we choose a different basis on $T_x \cs$, we might get
an additional factor. In fact,
the transformation matrix intertwining two bases is some $\Gl(\drei)$ element,
whose determinant is precisely that factor. 
Note that this prefactor may change from point to point, if we consider general
local frames. In the case we are interested in, however, the tangent bundle will be
globally trivial, such that we may assume that each frame is 
globally defined and such a change of bases corresponds to multiplication by
some function on full $\cs$.

To finally arrive at the definition of Ashtekar fields, observe that
multiplication of any tensorial object with $(\det e)^{-\weight}$
gives the corresponding tensor density of weight $\weight$.

\bdf
The \df{Ashtekar field} $E$ to a frame $\base$ is the densitized
frame field
\bgl
E & := & \inv{\det \base} \, \base
\egl
of weight $1$. 
\edf
Observe that this definition depends on the choice of a basis on each $T_x\cs$.
If that basis is given by the image of the canonical basis on $\R^n$,
then $\det\base$ is, of course, $1$. 
If $\drei \neq 1$, then the frame can be reconstructed from the Ashtekar field.
Use 
\bgl
\det E 
 & = & \det\Bigl(\inv{\det \base} \, \base\Bigr)
 \breitrel= \Bigl(\inv{\det \base}\Bigr)^n \det \base
 \breitrel= (\det \base)^{1-n}
\egl\noindent
to obtain
\bgl
\base 
 & = & (\det \base) E
 \breitrel= (\det E)^{\inv{1-n}} E \,.
\egl

\subsection{Orthonormal Frames}
\label{subsect:ortho-frames}
\bdf
A frame is called 
\bunum
\item
\df{orthonormal} w.r.t.\ the metric $\metrik$ on $\cs$ iff it is
an isometry%
\footnote{Here, as throughout the paper, we have equipped $\R^\drei$ with the standard
Euclidean metric.};
\item
\df{oriented} iff it preserves the orientation.
\eunum
\edf
Again, one may specify an orthonormal frame 
by an orthonormal basis of $T_x\cs$.\ \cite{KoNo1}

Any frame 
defines a metric 
such that 
the frame is orthonormal w.r.t.\ that metric.
In fact, if
$\stdskalprod\cdot\cdot$ denotes the Euclidean scalar product, then
\bgl
\metrik(X,Y) & := & \stdskalprod{\base^{-1}(X)}{\base^{-1}(Y)} 
\qquad \text{for $X,Y \in T_x\cs$} 
\egl\noindent
defines a metric $\metrik$ on $\cs$, such that $\base$ is an isometry.
While frames determine a metric uniquely, 
a metric does not fix the orthonormal frame.
In fact, $\base$ and $\base'$ are isometries for $q$ iff
$\base' = \base \circ L_g$ for some $g \in \Ocf(\drei)$, where
$L_g$ denotes the left translation by $g$. 

The bundle $\Ocf_\metrik(\cs)$ of orthonormal frames can be defined completely analogously
to that of general frames; one only has to replace $\Gl(\drei)$ by
$\Ocf(\drei)$. At the same time, $\Ocf_\metrik(\cs)$ is the 
reduction of the structure group $\Gl(\drei)$ of the frame bundle to the structure group
$\Ocf(\drei)$. Similar arguments apply to the bundle $\Ocf_\metrik^+(\cs)$ of oriented
orthonormal frames having structure group $\SO(\drei)$.

\section{Ashtekar Connection}
\label{sect:ashtekar-conn}
The other part of the Ashtekar variables is a metric connection
on the tangent bundle -- or in an appropriate associated principal 
fibre bundle. In each case, we continue to be given some Riemannian
metric $\metrik$ on some oriented manifold $\cs$. 
As we need a vector-product like structure on $T\cs$,
the dimension of $\cs$ is now required to be $3$. 
Later, for the definition of 
the Weingarten mapping, we assume $\cs$ to be an embedded submanifold of
a four-dimensional Lorentzian manifold $(M,\viermetrik)$, whereas
the metric $\metrik$ on $\cs$ is induced by $\viermetrik$ to $\cs$.

\subsection{Induced Vector Product}
\label{subsect:modif-kreuz-prod}
First we transfer the vector product on $\R^3$ by 
means of the metric to $T\Sigma$.

\bdf
Let $\metrik$ be a metric on $\cs$, and let $X$, $Y$ be tangent vectors on $\cs$.

For any oriented $\metrik$-orthonormal frame $\base$ define
\bgl
\ivp X Y & := & \base\bigl[\base^{-1}(X) \kreuz \base^{-1}(Y)\bigr]\,,
\egl
with $\kreuz$ being the standard vector product on $\R^3$.
\edf

\blem
$\ivpalone$ is well defined as 
it depends on $\metrik$ only, but not on the oriented 
$\metrik$-orthonormal frame $\base$ itself. 
\elem
\bpf
If $\invbase,\invanotherbase$ are inverses of oriented $\metrik$-orthonormal frames,
then there is an $A \in \SO(3)$ with 
$\invanotherbase(X) = A\invbase(X)$ for all $X$. Hence,
by the $\SO(3)$ invariance of the vector product, we have
\bgl 
\invanotherbase\bigl[\ivp[\invbase]XY\bigr] 
 & = & A\invbase\bigl[\ivp[\invbase]XY\bigr] 
 \breitrel= A\invbase(X) \kreuz A\invbase(Y) \\
 & = & \invanotherbase(X) \kreuz \invanotherbase(Y)  
 \breitrel= \invanotherbase\bigl[\ivp[\invanotherbase]XY\bigr] \,.
\egl
\qed
\epf
One immediately checks that the definition extends to a smooth
operation on vector fields, again denoted by $\ivpalone$.
The relevant properties of the vector product can easily be transferred to the induced
vector product:
\bprop
\label{prop:ivp-properties}
For all vector fields $X, Y, Z$, we have 
\bgl
\ivp X Y & = & - \ivp Y X \s
\skalprod{\ivp X Y}Z & = & \skalprod X{\ivp Y Z} \s
\ivp X{(\ivp Y Z)} & = & \skalprod X Z Y - \skalprod X Y Z \s
\egl
and the Jacobi identity
\bgl
\ivp X{(\ivp Y Z)} + \ivp Y{(\ivp Z X)} + \ivp Z{(\ivp X Y)}  
 & = & 0 \,.
\egl
\eprop
\bpf
Straightforward.
\qed
\epf

\brem
\bnum2
\item
The same product can be defined using the standard wedge product and
the Hodge operator w.r.t.\ $\metrik$. In fact,
$\ivp X Y = \hodge(X \keil Y)$.
\item
Note that, by assumption, $\cs$ is parallelizable, whence even globally
defined frames exist. 
\enum
\erem

\subsection{Weingarten Mapping}
\label{subsect:weingarten}
The second ingredient of the Ashtekar connection, 
the \df{Weingarten mapping}, corresponds to the second fundamental form
(also called exterior curvature by physicists).
It is defined by
\fktdefabgesetzt{\wgabb}{T\cs}{T\cs\,.}{X}{\viernabla_X n}
\noindent
Here, $n$ is the normal to $\cs$ within $(M,\viermetrik)$,
and $\viernabla$ is the Levi-Civita connection for $\viermetrik$ on $M$.
By metricity of $\viernabla$, the Weingarten mapping is well defined.%
\footnote{We have $\wgabb(X) \in T\cs$, because
$\skalprod{\wgabb(X)}n \ident \skalprod{\viernabla_X n}n = \inv2 X \skalprod n n = 0$.}
Moreover it is symmetric%
\footnote{See $\skalprod{\wgabb(X)}Y 
  \ident  \skalprod{\viernabla_X n}Y 
  =  - \skalprod{n}{\viernabla_X Y} 
  =  - \skalprod{n}{\viernabla_Y X} + \skalprod{n}{[Y,X]} 
  =  \skalprod{\viernabla_Y n}X 
  \ident  \skalprod{\wgabb(Y)}X$\,,
using metricity and torsion-freeness of the Levi-Civita connection.
}
and $C^\infty(\cs)$-linear.
The second fundamental form $\extcurv$ 
is given by $\extcurv(X,Y) = \skalprod{\wgabb(X)}Y$.
The other way round, this equation allows to obtain $\wgabb$ from $\extcurv$
as $\metrik$ is non-degenerate.

\subsection{Ashtekar Connection}
\label{subsect:ashtekar-conn}
\bdf
The \df{Ashtekar connection} w.r.t.\ \df{Barbero-Immirzi parameter} $\bip \in \C$
is defined by
\bgl
\ashkov_X Y & := & \nabla_X Y + \bip\,\ivp{\wgabb(X)}Y \,.
\egl
\edf
We consider only the case $\bip \neq 0$.
Note that the imaginary part of $\bip$ may be non-zero. 
Indeed, originally, $\bip$ has been set
to $\I$ as only then some unpleasant term within the Hamiltonian constraint,
prefactored by $(1+\bip^2)$, disappeared. Of course, if $\bip$ is not real,
then, in what follows, 
the connections etc.\ will live in the corresponding complexified structures.
If necessary, we will tacitly assume this (without further indication).
\bprop
$\ashkov$ is metric and obeys the Leibniz rule
\bgl
\ashkov_X (\ivp Y Z)
 & = & \ivp{\ashkov_X Y}Z + \ivp Y{\ashkov_X Z}
\egl
\eprop
\bpf
\bnum2
\item
The metricity of the Levi-Civita connection just reads as 
\bglklein
X\skalprod Y Z & = & \skalprod{\nabla_X Y}Z + \skalprod Y{\nabla_X Z} \\
\eglklein
for all vector fields $X$, $Y$, $Z$.
Additionally, 
\bglklein
0 & = & \skalprod{\ivp{\wgabb(X)}Y}Z + \skalprod {\ivp Y{\wgabb(X)}} Z \\
  & = & \skalprod{\ivp{\wgabb(X)}Y}Z + \skalprod Y{\ivp{\wgabb(X)} Z}
\eglklein%
by Proposition \ref{prop:ivp-properties}.
Adding this to the equation above, we get
\bglklein
X\skalprod Y Z & = & \skalprod{\ashkov_X Y}Z + \skalprod Y{\ashkov_X Z} \,,
\eglklein
hence metricity of the Ashtekar connection.
\item
To prove the Leibniz rule, add the Leibniz rule 
\bglklein
\nabla_X(\ivp Y Z) & = & \ivp{\nabla_X Y} Z + \ivp Y {\nabla_X Z} \\
\eglklein
for the Levi-Civita connection (see Appendix \ref{app:leibniz})
to 
\bglklein
\ivp{\wgabb(X)}{\bigl(\ivp Y Z\bigr)} & = & \ivp{\bigl(\ivp{\wgabb(X)}Y\bigr)} Z + \ivp Y{\bigl(\ivp{\wgabb(X)}Z\bigr)}
\eglklein%
which is immediately derived from Jacobi's identity and the antisymmetry of $\ivpalone$.
\qed
\enum
\epf

\neueseite

\subsection{Reconstruction}
It is possible to reconstruct 
both the metric $\metrik$ and the second 
fundamental form $\extcurv$ from $E$ and $\ashkov$.
In fact, first reconstruct $\metrik$ from $E$ as in Section~\ref{sect:frames}.
Then, this metric $\metrik$ determines the Levi-Civita connection $\nabla$. From this, 
as $\bip \neq 0$, we may
regain $\ivp{\wgabb(X)}Y$ for all vector fields $X,Y \in T\cs$.
Since $2 \stdx = \sum_{i} \stdbasis_i \kreuz (\stdx \kreuz \stdbasis_i)$
for the standard orthonormal 
basis $(\stdbasis_1,\stdbasis_2,\stdbasis_3)$ of $\R^3$,
we get
\bgl
\wgabb(X) 
 & = & \inv2 \sum_{i=1}^3 
       \ivp{\base(\stdbasis_i)}{\bigl(\ivp{\wgabb(X)}{\base(\stdbasis_i)}\bigr)} \,.
\egl\noindent
As already mentioned, the second fundamental form can now easily be derived from this
$\wgabb$.

\subsection{Curvature}

\bprop
\label{prop:tors+curv-ash}
The torsion of the Ashtekar connection is given by
\bgl
T^A(X,Y) 
 & = & \bip\,\bigl[\ivp{\wgabb(X)}Y - \ivp{\wgabb(Y)}X\bigr]
\egl
and the curvature by
\bgl
R^A(X,Y)Z 
  & = & R(X,Y)Z + \bip^{\phantom2}\ivp{[(\nabla_X \wgabb) Y - (\nabla_Y \wgabb) X]}Z
        \\&&\phantom{R(X,Y)Z}+ \bip^2 \ivp{[\ivp{\wgabb(X)}{\wgabb(Y)}]}Z

\egl
\eprop
\bpf
\bnum2
\item
The torsion of the Ashtekar connection is given by
\bgl
T^A(X,Y) 
 & = & \ashkov_X Y - \ashkov_Y X - [X,Y] \\
 & = & \nabla_X Y - \nabla_Y X - [X,Y] + \bip\,\ivp{\wgabb(X)}Y - \bip\,\ivp{\wgabb(Y)}X\\
 & = & \bip\,\ivp{\wgabb(X)}Y - \bip\,\ivp{\wgabb(Y)}X
\egl
as the Levi-Civita connection is torsion-free.
\item
For the curvature observe (with $\wgabb$ rescaled by $\bip\wgabb$)
\bgl
R^A(X,Y)Z 
  & = & \ashkov_X \ashkov_Y Z - \ashkov_Y \ashkov_X Z - \ashkov_{[X,Y]} Z \\
  & = & \nabla_X \nabla_Y Z - \nabla_Y \nabla_X Z - \nabla_{[X,Y]} Z
         \\&&{}
         + \nabla_X \bigl(\ivp{\wgabb(Y)}Z\bigr) 
         - \ivp{\wgabb(Y)}{\nabla_X Z} 
         \\&&{}
	  - \nabla_Y \bigl(\ivp{\wgabb(X)}Z\bigr) 
         + \ivp{\wgabb(X)}{\nabla_Y Z} 
         \\&&{}
	 - \ivp{\wgabb([X,Y])}Z 
	 \\&&{}+ \ivp{\wgabb(X)}{\bigl(\ivp{\wgabb(Y)}Z\bigr)}
         - \ivp{\wgabb(Y)}{\bigl(\ivp{\wgabb(X)}Z\bigr)} \\
  & = & \nabla_X \nabla_Y Z - \nabla_Y \nabla_X Z - \nabla_{[X,Y]} Z
         \\&&{}
         + \ivp{\nabla_X (\wgabb(Y))}Z 
	 - \ivp{\nabla_Y(\wgabb(X))}Z
         \\&&{}
	 - \ivp{\wgabb(\nabla_X Y - \nabla_Y X)}Z 
	 \\&&{}+ \ivp{\wgabb(X)}{\bigl(\ivp{\wgabb(Y)}Z\bigr)}
         + \ivp{\wgabb(Y)}{\bigl(\ivp Z{\wgabb(X)}\bigr)} \\
  & = & R(X,Y)Z + \ivp{[(\nabla_X \wgabb) Y - (\nabla_Y \wgabb) X]}Z
        \\&&{}+ \ivp{[\ivp{\wgabb(X)}{\wgabb(Y)}]}Z \,.
\egl
\qed
\enum
\epf\noindent
A few of the symmetries of the Riemann curvature tensor survive the 
transition to Ashtekar connections. As $\ashkov$ is a covariant derivative,
we have 
\bgl
R^A(X,Y) & = & -R^A(Y,X)\,,
\egl\noindent
and as $\ashkov$ is metric, we get
\bgl
\skalprod{R^A(X,Y)Z}W & = & -\skalprod{R^A(X,Y)W}Z \,.
\egl

\subsection{Frame Bundle Connection}
\label{subsect:frame-bundle-conn}
Taking the canonical action $\action$ of $\Gl(\drei)$ and its subgroups 
on $\R^\drei$, we may consider the tangent bundle as an associated bundle
of the frame bundle via the usual isomorphism
\bgl
\Gl(\cs) \kreuz_{(\Gl(\drei),\action)} \R^\drei  & \iso & T\cs \,.\\\
[\base,\stdx] & \auf & \base(\stdx)  
\egl\noindent
This isomorphism induces an isomorphism between the 
space of covariant derivatives on $T\cs$ and 
the space of connection $1$-forms in the frame bundle.
Similarly, given any metric $\metrik$ on $\cs$, we have 
\bgl
\Ocf_\metrik(\cs) \kreuz_{(\Ocf(\drei),\action)} \R^\drei
\breitrel\iso
\Ocf_\metrik^+(\cs) \kreuz_{(\SO(\drei),\action)} \R^\drei
& \iso & T\cs 
\egl\noindent
for the (oriented) orthonormal frame bundle,
and an isomorphism between the metric covariant derivatives
on $T\cs$ and connections in $\Ocf_\metrik(\cs)$.
Therefore, we can now naturally transfer the Ashtekar connection
from $T\cs$ to the frame bundles.
Given a local frame $\base$ on $U \teilmenge \cs$, we may define
a local connection $1$-form 
\bgl
\laz & := & \skalprod{\ashkov\base}\base
 \breitrel: TU \nach \gl(3) \,.
\egl\noindent
A bit more explicitly, for each vector field $X$, we get a
$\gl(3)$ matrix $A(X)$ by
\bgl
\stdskalprod{\laz(X) \stdx}{\stdy} 
 & = & \skalprod{\ashkov_X[\base(\stdx)]}{\base(\stdy)}
\egl\noindent
for all $\stdx,\stdy\in\R^3$, viewing both sides as functions on $U$.
The Ashtekar connection $\ashconn$ in the frame bundle is now the connection
that is mapped to $\laz$ by the different pull-backs 
$\anotherbase^\ast \ashconn$.
Here, we viewed the frame $\anotherbase$ as
a mapping from some $U$ to $\Gl(\cs)$.
If we assume $\base$ to be $\metrik$-orthonormal,
then $\ashconn$ is even a connection in the orthonormal frame bundle $\Ocf_\metrik(\cs)$,
and its local versions $\laz$ are $\Ocf(3)$ connections.
Indeed, if $\base$ is orthonormal,
then for fixed $\stdx,\stdy\in\R^3$
\bgl
\stdskalprod{\laz(X) \stdx}{\stdy}
 & = & \skalprod{\ashkov_X[\base(\stdx)]}{\base(\stdy)} \\
 & = & \ashkov_X \skalprod{\base(\stdx)}{\base(\stdy)}
          - \skalprod{\base(\stdx)}{\ashkov_X[\base(\stdy)]} \erl{by metricity of $\ashkov$} \\
 & = & \ashkov_X \stdskalprod{\stdx}{\stdy}
          - \skalprod{\ashkov_X[\base(\stdy)]}{\base(\stdx)} \erl{$\base$ isometry} \\
 & = & - \stdskalprod{\stdx}{\laz(X) \stdy} \,.
\egl\noindent
Similarly, in the oriented case, we get $\SO(3)$ connections.
From the definitions, we get
\bgl
\ashconn & = & \lcconn + \bip\,\bdlextcurv
\egl\noindent
with 
\bgl
\bdlextcurv(X) & = & \skalprod{\wgabb(\pi_\ast X)}{\ivp \base \base}
\egl\noindent
for all vector fields $X$ on the frame bundle, $\pi$ being the 
bundle projection.

\subsection{Physics Notation}
\label{subsect:index-version}
In the physics literature, Ashtekar connections are given by their
components in some coordinate system. 
To reproduce this, let $\{M_1,M_2,M_3\}$ be a basis of $\so(3)$ with
$[M_i,M_j] = \varepsilon_{ij}{}^k M_k$, and let $\chart : U \nach \R^3$ 
be some chart for open $U \teilmenge \cs$.
Then we get a local basis $\{\del_1,\del_2,\del_3\}$ for the
tangent space. Finally, choose some orthonormal frame $\base$,
viewed as a map from $U$ to $\Ocf_\metrik^+(\cs)$.
Hiding the dependence on $U$, we get 
$\base(\stdbasis_i) =: \base_i^a \del_a$, with $\stdbasis_i$ being
the $i$-th vector in the standard basis of $\R^3$.
It is now a bit lengthy, but straightforward \cite{philipp-dipl} to
show that 
\bgl
\base^\ast \lcconn(\del_a)
 & = & \inv2  \varepsilon_{ij}{}^k \,\skalprod{\nabla_{\del_a}\base(\stdbasis_i)}{\base(\stdbasis_j)} M_k
 \breitrel= \Gamma_a^k M_k
\egl\noindent
reproduces the quantity $\Gamma_a^k$ that often is called spin connection (whose geometric
definition will be given in Section \ref{sect:spin}).
Similarly, one checks \cite{philipp-dipl} that
\bgl
\base^\ast \bdlextcurv(\del_a)
 & = & \varepsilon^{ijk} \skalprod{\wgabb(\del_a)}{\base(\stdbasis_i)} [M_j,M_k]
 \breitrel= k(\del_a,\del_b) \base^{b}_i M^i
 \breitrel= k_{a}^i M_i
\egl\noindent
reproduces the extrinsic curvature $k_a^i$.
Altogether, we have
\bgl
\base^\ast \ashconn(\del_a) 
 & = & (\Gamma_a^i + \bip k_a^i) M_i \,.
\egl\noindent
As an $\so(3)$ Lie algebra basis can always be regarded also as an $\su(2)$ Lie algebra
basis, the expression above is also a local coordinate version of the $\SU(2)$
Ashtekar connection. The deeper geometric meaning of this transition is
encoded in a spin structure to be considered below.

\section{Spin Structures}
\label{sect:spin}
\bdf
A \df{spin structure} \cite{Friedrich} on $(\cs,\metrik)$ is a pair
$(\spinbdl(\cs),\coveroben)$ consisting of 
\bunum
\item
an $\SU(2)$ principal fibre bundle $\widetilde\pi : \spinbdl(\cs) \nach \cs$ and
\item
a double covering $\coveroben : \spinbdl(\cs) \nach \Ocf_\metrik^+(\cs,\metrik)$,
\eunum
such that the following diagram commutes:
\newcommand{\stddiagram}[5][]{
{\begin{center}\begin{minipage}{#4}%
                            \begin{diagram}[labelstyle=\scriptstyle,height=\CDhoehe,l>=\CDstdlaenge,#1]%
                            #5%
                            \end{diagram}\end{minipage}\end{center}}}
\stddiagram[height=1.3em]{6}{-2}{14em}{
 S(\Sigma) \kreuz \SU(2) & \relax\rnach & S(\Sigma) \\
 &&& \rdnach^{\widetilde\pi} \\
 \relax\dnach^{\text{both } 2:1}_{\coveroben \kreuz \covergroup} & & \relax\dnach^{2:1}_\coveroben & & \Sigma \,.\\ 
 &&& \runach^{\pi} \\
 \Ocf_\metrik^+(\cs) \kreuz \SO(3) & \relax\rnach & \Ocf_\metrik^+(\cs) 
}
The horizontal lines denote the action of the respective structure group,
and $\covergroup$ the double cover $\covergroup : \SU(2) \iso \Spin(3) \nach \SO(3)$.
\edf
As $\covergroup_\ast : \su(2) \nach \so(3)$ is an isomorphism, any
connection $\conn$ in $\Ocf_\metrik^+(\cs)$ can be lifted to a connection 
$\widetilde\conn$ via
\bgl
\covergroup_\ast \circ \widetilde\conn
 &  := & \coveroben^\ast \conn \,.
\egl\noindent
The spin bundle connection $\spinconn$ corresponding to the 
Levi-Civita connection is called \df{spin connection},
and the spin bundle connection $\ashconnspin$ 
corresponding to the $\SO(3)$ Ashtekar connection $\ashconn$ is the
celebrated $\SU(2)$ Ashtekar connection.
Completely analogously to the case of frame bundles, the tangent bundle is
associated to the spin bundle
via 
\bgl
T\cs & \iso & \spinbdl(\cs) \kreuz_{(\SU(2),\action\circ\covergroup)} \R^3 \,.
\egl\noindent
Hence, if we assumed connections in a principal fibre bundle 
to be the fundamental ones, for the Ashtekar connection
acting on the tangent bundle it did not matter, 
whether we had derived it from the $\SU(2)$ or the $\SO(3)$
version.

\section{Application to Cosmology}
\label{sect:cosmol}
Let us finally discuss highly symmetric models as used in quantum cosmology.
Here, we would like to restrict ourselves to Friedmann-Robertson-Walker models.
In other words, we assume that the spacetime manifold $M$
is split as $M = I \kreuz \cs$, where $I$ is some open interval in $\R$,
and the spacetime metric is given by
\bgl
g & = & -\dd \tau^2 + a^2 q \,.
\egl\noindent
Here, $q$ is a metric on $\cs$ of constant sectional curvature $\kappa$,
\pagebreak
and $a : I \nach \R$ is the scale factor.
Observe that the normal $n$ to the Cauchy slice equals $\del_\tau$. 
Using the Koszul formula 
with $[X,Y] = [Y,n] = [n,X] = 0$,
we get
\bglklein[0.4ex]
2\skalprod{\wgabb(X)}Y 
 & \ident & 2\skalprod{\viernabla_X n}Y 
 \breitrel= X \skalprod n Y - Y \skalprod n X + n \skalprod X Y  \s
 & = & \del_\tau [a^2 q(X,Y)] 
 \breitrel= 2 \dot a a q(X,Y) 
 \breitrel= 2 \skalprod{\frac{\dot a}a X} Y \,, \\
\eglklein%
\noindent
hence 
\bgl
W & = & \hubble \: \EINS
\egl\noindent
with the Hubble ``constant'' 
\bgl
\hubble & := & \frac{\dot a}a \,.
\egl\noindent
Thus, the Ashtekar connection has a very simple form
\bgl
\ashkov_X Y \breitrel{=} \nabla_X Y + \bip \hubble \: \ivp{X}Y  \,.
\egl\noindent
Its torsion and curvature 
are given by
\bgl
T^A(X,Y) & = & 2\bip \hubble\: \ivp X Y \\
R^A(X,Y)Z & = & \bigl[(\bip \hubble)^2 - \kappa\bigr]\ivp{(\ivp X Y)}Z 
\egl\noindent
In fact, the torsion formula follows immediately from 
Proposition \ref{prop:tors+curv-ash}. For the second one, observe additionally that
a space of constant sectional curvature $\kappa$ fulfills
\cite{KoNo1} 
\bgl
R(X, Y )Z
 & = & \kappa \bigl(\skalprod Z Y X - \skalprod Z X Y \bigr) 
 \breitrel= \kappa \: \ivp Z{(\ivp X Y)} \,.
\egl\noindent

\appendix
\section{Leibniz Rule for Levi-Civita Connection}
\label{app:leibniz}
In this appendix we are going to show that for all $X,Y,Z \in T\cs$ the Leibniz rule
\bglklein
\label{eq:leibniz-lc-allg}
\nabla_X(\ivp Y Z) & = & \ivp{\nabla_X Y} Z + \ivp Y {\nabla_X Z} 
\eglklein\noindent
holds for the Levi-Civita connection. 
For this, let $(\stdbasis_1,\stdbasis_2,\stdbasis_3)$ be the standard basis of $\R^3$
and let $\base$ be some orthonormal frame. Define $\base_i := \base(\stdbasis_i)$ 
for $i = 1,\ldots,3$.
Then we have to show that 
for all $j,k,l$ and $X$
\begin{eqnarray}
\nonumber
 &   &  \hspace*{-8em}\skalprod{\nabla_X {\base_l}}{\ivp {\base_j}{\base_k}}
           + \skalprod{\nabla_X {\base_j}}{\ivp {\base_k}{\base_l}}
           + \skalprod{\nabla_X {\base_k}}{\ivp {\base_l}{\base_j}} \\
 & = &  - \skalprod{\nabla_X (\ivp {\base_j}{\base_k})}{{\base_l}}
           + \skalprod{\ivp{\nabla_X \base_j} \base_k}{{\base_l}}
           + \skalprod{\ivp{\base_j}{\nabla_X \base_k}}{{\base_l}}
 \breitrel= 0	\,.    \hspace*{-4em}
\label{eq:leibniz-lc-spez}
\end{eqnarray}
In fact, first observe%
\footnote{Note that we do not use summation convention here.} 
that
\bgl
\skalprod{\nabla_X \base_i}{\base_i} 
 & = & \nabla_X \skalprod{\base_i}{\base_i} - \skalprod{\base_i}{\nabla_X \base_i} 
 \breitrel= - \skalprod{\nabla_X \base_i}{\base_i} 
\egl\noindent
for $X \in T\cs$, since 
$\nabla$ is metric, $\metrik$ is symmetric, and
$\skalprod{\base_i}{\base_i} \ident 1$; consequently,
\begin{eqnarray}
\label{eq:nabla-i-i-0}
\skalprod{\nabla_X \base_i}{\base_i} & = & 0 \,.
\end{eqnarray}
Then we consider the four relevant cases:
\bnum4
\item
If $j,k,l$ are pairwise different, then the right-hand sides of the scalar products 
in the upper line of \eqref{eq:leibniz-lc-spez} 
are always plus/minus the basis element at the respective
left-hand side, whence each scalar product vanishes by \eqref{eq:nabla-i-i-0}.
\item
If $j = k$, then the second and the third scalar product
cancel each other due to the antisymmetry of $\ivp{}{}$.
For the same reason, the first scalar product in 
\eqref{eq:leibniz-lc-spez} vanishes. 
\item
If $j = l$, then the first and the second scalar product cancel; the
third one vanishes.
\item
If $k = l$, then the first and the third scalar product cancel; the 
second one vanishes.

\enum

\neueseite

\section*{Acknowledgments}
The authors thank Johannes Brunnemann, Stefan Funkner, 
S\"oren Graupner and, in particular, Thomas Leistner for interesting discussions 
as well as Maximilian Hanusch and Matthias Schmidt for
their helpful comments on a draft version of this article.
One of the authors [CF] has been supported by the Emmy-Noether-Programm of
the Deutsche Forschungsgemeinschaft under grant FL~622/1-1.

\end{document}